\documentstyle[11pt,paspconf]{article}
\input psfig.sty

\def\DF{DF}

\begin{document}

\title{Galaxy dynamics by $N$-body simulation}
\author{J. A. Sellwood}
\affil{Rutgers University, Department of Physics \& Astronomy, PO Box 849,
Piscataway, NJ 08855}

\bigskip {\parindent=1.1cm \hangindent=\parindent
Rutgers Astrophysics Preprint No 206

To appear in ``Computational Astrophysics'' edited by D. A. Clarke \&

M. J. West, ASP Conference Series}

\begin{abstract}
I compare various popular and unpopular techniques for simulating large
collisionless stellar systems.  I give a quantitative comparison of the raw
cpu times required for five separate codes, including tree codes and basis
function expansions, which demonstrate that grid codes are most efficient for
large numbers of particles.  Since efficiency is only one consideration when
choosing a code, I discuss other strengths and weaknesses of the various
methods.  While some applications may require the maximum possible number of
particles, I argue that quiet start techniques can often permit reliable
results to be obtained with moderate particle numbers.  I suggest a quiet
start procedure for spherical stellar systems and show that it leads to a
significant reduction in the relaxation rate.  The combination of efficient
codes and quiet starts allows many galactic dynamical problems to be tackled
without the need for supercomputers.
\end{abstract}

\section{Introduction}

In this very brief review, I discuss several techniques that attempt to mimic
the gravitational dynamics of galaxies using $N$-body methods.  The
collisionless property of these stellar systems demands algorithms that try,
as far as possible, to suppress the relaxation caused by $\sqrt{N}$-type
fluctuations in the particle distribution.  The objective therefore is quite
different from collisional problems, discussed by Hut (these proceedings), in
which the interactions between every pair of particles have to be treated as
accurately as possible.

My purpose is to emphasize two quite old, but very powerful, methods which I
feel are not being fully exploited in much of the current work in the field.
The first is the superior efficiency of particle-mesh (PM) techniques over
all others that are widely used, which I demonstrate by presenting some
timing comparisons of five codes.  While efficiency is important, it is not
the only factor that determines the choice of code, and I discuss the
strengths and weaknesses of each of these methods.  Secondly, I wish to
underline the advantage of using a quiet start, which I show reduces
noise-driven relaxation for a fixed number of particles.

For all the tests reported here, I used a stellar polytrope of index 5, or
Plummer sphere, which has the density profile $$
\rho(r) = {3M \over 4\pi a^3} \left( 1 + {r^2 \over a^2} \right)^{-5/2},
\eqno(1)
$$ where $M$ is the total mass and $a$ is a length scale.  A spherical model
with an isotropic distribution function of this form is stable (Binney \&
Tremaine 1987).  This mass distribution is not very centrally condensed,
making it particularly easy to simulate.  This infinite mass distribution has
to be truncated to fit inside a grid; for all these tests, I therefore
restrict the spatial extent of the particles to $r<6a$ at the outset, thereby
discarding just over 4\% of the mass.

\begin{figure}[t]
\centerline{\psfig{figure=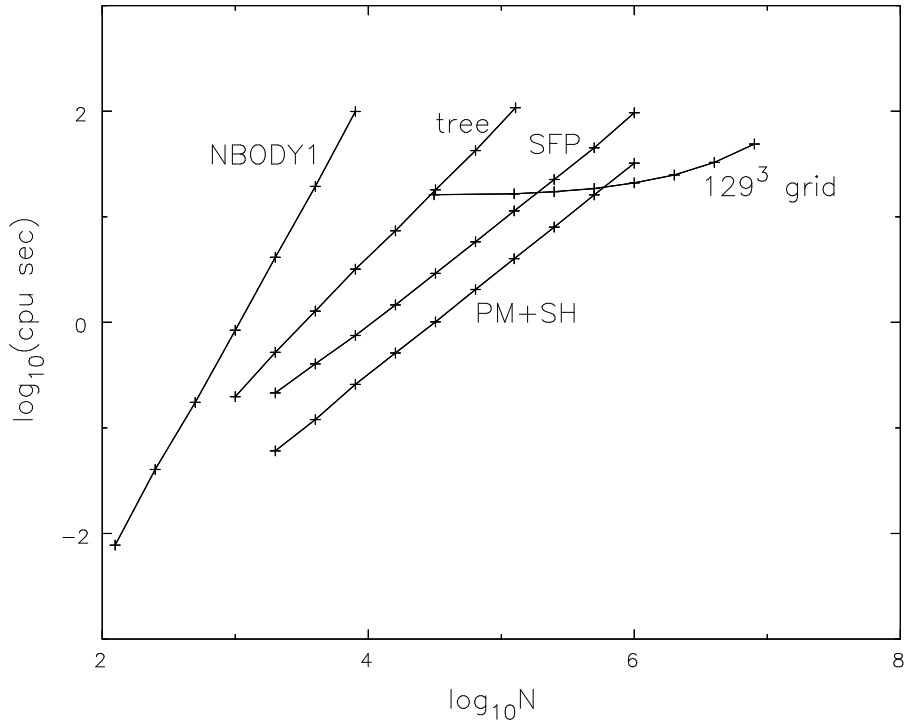,width=0.90\hsize,clip=,angle=270}}
\kern -30pt
\caption{Comparison of the performance of five codes for collisionless
$N$-body simulations.  The lines join values obtained from sequences of runs
with different numbers of particles.  While the spatial resolution is kept
fixed for each method, it differs from method to method.}
\end{figure}

\section{Performance tests}
Figure 1 compares the cpu time required to complete one time step using five
different methods.  Timings are all on a DEC alphastation 500/333.

The methods in this comparison are two PP (particle-particle) methods, direct
and tree, a basis function expansion method, or SFP (smooth field-particle)
and a generic PM (particle-mesh) code.  The fifth method is based on an
expansion in surface harmonics, which has been widely used in the past, and
is a hybrid PM+SH.  Of these, the SFP, PM and PM+SH codes are much easier to
vectorize/parallelize than are the PP methods.

\subsection{Direct method}
The simplest method of all is NBODY1, the least sophisticated of the sequence
of direct $N$-body codes devised by Aarseth (1985); the Fortran source is
reproduced in an appendix of Binney \& Tremaine (1987).  I set the softening
length $\epsilon = 0.01a$ and accuracy parameter $\eta = 0.03$.  Since the
individual time steps for each particle are determined by the adopted
accuracy criterion, I have defined the ``time per step'' as one twentieth of
the cpu time to reach one dynamical time $=(a^3/GM)^{1/2}$, which is the time
step length I used for all the other codes.

A least-squares fit to the points in this log-log plot has a slope of 2.3,
i.e, steeper than the usually quoted ${\cal O}(N^2)$ behavior.  The steeper
dependence is because the number of close neighbors rises with $N$, which
implies that a larger fraction of the particles requires shorter steps, even
though the mass of each is correspondingly less.  Performance could, of
course, be improved by increasing the softening parameter or relaxing the
accuracy criterion.  One could justifiably argue that the softening length
should be reduced as $N$ is increased, a strategy that would steepen the $N$
dependence still more.

The rapid increase of cpu time needed with $N$ renders this code far too
expensive for desirable numbers of particles and more efficient techniques
should be used.  The GRAPE machines implement an algorithm that is not very
different; their strategy seems essential for collisional problems which
require highly accurate forces, but collisionless problems allow other
algorithms to be employed which yield results more efficiently -- more than
compensating for the advantage of custom built chips.

\subsection{Tree code}

The line marked ``tree'' indicates timings obtained using Hernquist's (1987)
public-domain tree code, which employs the original Barnes \& Hut (1986)
algorithm.  Admittedly this may no longer be regarded as a state-of-the-art
code, but most recent improvements have been directed towards improving
accuracy while running speeds do not appear to have improved significantly.
In fact, the speeds reported by Dubinski (1996) for an $N = 640$~K
calculation on the T3D using 128 nodes are only some $\sim 20$ times higher
than the extrapolated line in Figure~1, which is for a single processor
workstation.

The timings reported in Figure 1 are with quadrupole forces included but use
the large opening angle $\theta=1$, which leads to quite low accuracy.  A
more accurate calculation with a smaller opening angle would, of course, take
longer.  The slope of the line shown in Figure 1 is approximately 1.3, which
is steeper than the predicted ${\cal O}(N\log N)$.

\subsection{Surface harmonic method}

The PM+SH method I employ has not been described elsewhere.  It sits between
the fully particle-based approach used by Villumsen (1982), White (1983) and
McGlynn (1984) and the fully grid-based method used by van Albada (1982).  I
tabulate the set of active coefficients for the expansion of the potential at
a number of fixed radii (a mesh) and derive the acceleration applied to each
particle from linearly interpolated values of these coefficients at the
radius of the particle.  There is no need to sort particles and there is no
gridding in the non-radial directions.  The radii of grid points can be more
closely spaced near the center and there should be many particles between
successive grid points.

With this scheme, the cpu time rises linearly with $N$ and is virtually
independent of the number of radial grid points for fixed $l_{\rm max}$.  The
timings shown in Figure 1 are for $l_{\rm max}=6$ with 201 radial grid points.

\subsection{Basis function expansion}

The line marked SFP in Figure 1 is for a basis function expansion method.  It
was obtained using Hernquist's so-called SCF code, but I prefer the acronym
SFP (smooth field-particle), since all $N$-body codes have self-consistent
fields.  Hernquist \& Ostriker (1992) derive a new basis, but otherwise their
algorithm is identically that first proposed by Clutton-Brock (1972, 1973).

As advertized, the slope of the line in this plot is precisely unity when the
number of functions is held fixed.  The data in Figure 1 were obtained using
the Clutton-Brock (1973) basis, employing radial functions $0\leq n\leq10$
and an angular expansion up to $l_{\rm max}=6$.  One
could reasonably argue that more functions should be employed as $N$ rises,
which would steepen the slope.  Unlike the method described in \S2.3, this
method becomes more time consuming as the radial resolution is increased,
since more functions need to be evaluated.

At the end of his paper, Clutton-Brock (1972) conceded that the SFP method
was not competitive, in terms of raw speed, with the Fourier grid methods
which were then emerging, a conclusion that has not changed in the subsequent
almost two and a half decades.

\begin{figure}[t]
\centerline{\psfig{figure=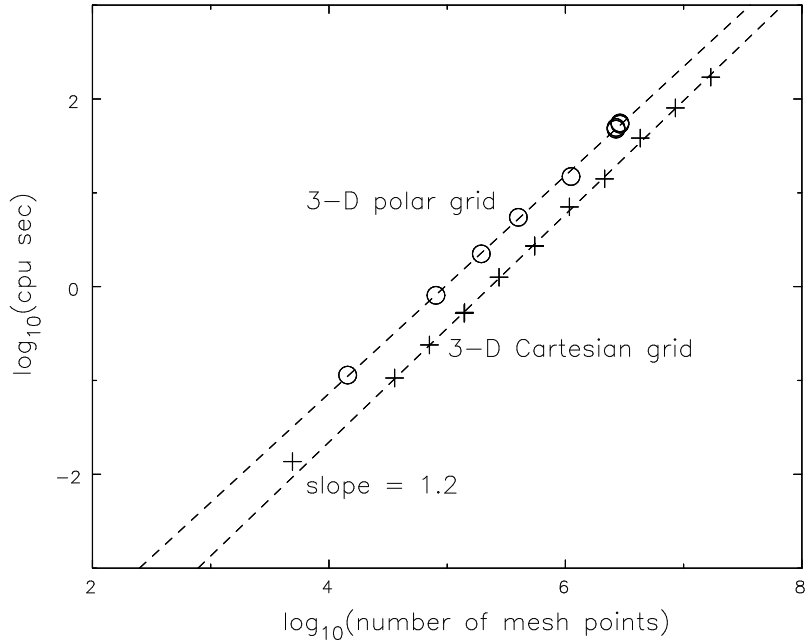,width=0.8\hsize,angle=270}}
\kern -30pt
\caption{The cpu time requirements for the field determination on two
different grids for various numbers of grid cells.  These times do not
include any particle movement.}
\end{figure}

\subsection{Grid methods}

The curve marked ``$129^3$ grid'' shown in Figure 1 results from a Cartesian
PM code with this number of grid cells.  It incorporates James's (1977)
Poisson solver which represents the only significant algorithmic improvement
since the first 3-D grid results reported by Hockney \& Brownrigg (1974).
It is clear from the near horizontal portion of the line at small $N$ that
the calculation time is dominated by the field determination except for very
large $N$.  In fact the slope is still well below linear at the last point
which is for $N=8$M.  The dominance of the field calculation part leads to
similar $N$-dependence for grids of other geometries.

Figure 2 illustrates how the $N$-independent field calculation time varies
with the size of grid employed.  Timings are both for James's Poisson solver
and my own 3-D cylindrical polar grid (Sellwood \& Valluri 1997), which
differs only in details from that described by Pfenniger \& Friedli (1993).
The potential returned on the Cartesian grid has the full resolution in all
three dimensions afforded by each grid.  The forces determined on the polar
grid, on the other hand, have azimuthal harmonics restricted to $0 \leq m
\leq 8$ only.  The polar grid is clearly more time consuming, but has the
compensating advantage of offering higher spatial resolution near the
symmetry axis, where the density of particles is generally greatest.

Other grid geometries are clearly possible.  van Albada (1982) used a
spherical grid to follow mildly aspherical
collapses with spectacular radial resolution.  To have conducted such
simulations on a fixed Cartesian grid with any degree of validity would have
required an impossibly large number of cells.  The moral here is that when
resolution demands a huge grid, then a different grid geometry would probably
be more appropriate.

Adaptive mesh refinement has been shown at this meeting to be highly
successful for other applications.  Many adaptive PM codes have been
described in the recent literature, most of them devised for the problem of
cosmological structure formation in which many dense regions develop at
random locations on the main grid; they seem ideally suited to the intended
application (e.g.\ Couchman, these proceedings).  I have not yet found a need
for such a method to follow the
global dynamics of an isolated galaxy, preferring instead to tailor the grid
to the mass distribution under study.  Of course, an adaptive refinement
strategy on a Cartesian grid would be immensely superior to a fixed grid for
the collapse problems studied by van Albada, yet his well-chosen spherical
grid required no adaptivity and remarkably few mesh cells.

\section{Which method to use?}

While raw cpu speed is clearly an important factor when determining which
code to employ for a problem, the physical properties of the model to be
studied may render a less efficient code more appropriate.

The principal disadvantage of Eulerian PM codes is that the fixed volume and
geometry of the mesh make them unsuited to following wholesale rearrangements
of the mass distribution, as occur during major galaxy mergers, for example.
This same inflexibility means that different grids are required for different
problems.  As these codes are already more complex than PP codes, the need to
rewrite for a new grid is a further significant handicap.  Furthermore, grid
codes generally require a few hundred MB of memory in addition to that needed
to store the particle coordinates.

Finally, choosing an interpolation scheme that will best hide the discrete
nature of the grid is a ``black art''; fortunately this problem has been
studied in great detail by the plasma physicists, who employ similar codes,
and the theory behind the various strategies has been developed at some
length; see e.g., Birdsall \& Langdon (1991).  Nevertheless, it is not
possible to hide the grid completely, and there are some delicate problems
for which grid effects are intolerably large.  For example, my attempts to
study fully three dimensional warped disks have been compromised by the
tendency for a thin flat disk inclined at an angle to the grid planes to
experience a weak torque from the grid that causes it to precess and to try
to align with the grid planes.  Until this tendency can be effectively
removed, a grid method simply cannot be used for this problem.  Fortunately,
serious problems of this kind are rare.

A common, but generally misdirected, criticism of grid methods is that they
lack spatial resolution.  It has to be admitted that Cartesian grids cannot
handle steep density gradients, but other grid geometries can (c.f. van
Albada 1982 and the PM+SH code described in \S2.3).  Furthermore, should
problems be identified for which a single fixed grid cannot resolve high
density regions, one could resort to adaptive mesh refinement, at the cost of
some extra software effort.

The criticism that current grid methods are inflexible and cannot follow
major rearrangements of the mass distribution can be applied {\it a
fortiori\/} to SFP methods.  Weinberg's (1996) extension of the SFP technique
to a dynamically changing basis attempts to remedy this inflexibility. 
Nevertheless, the form of the potential that can be represented by the few
hundred terms employed in typical SFP implementations is much more restricted
than can be represented on a grid of some million separate mesh points. 
(See also \S4.1.)

Both surface harmonic methods, and SFP codes in which the non-radial part of
the basis uses an expansion in $Y_l^m$ functions, are ill-suited to mass
distributions, such as disks, that are far from spherically symmetric.

Basis function expansions are best suited to simulations in which only very
minor changes in the mass distribution occur.  In fact, Earn \& Sellwood
(1995) found the method to be ideal for following the linear growth of
instabilities, where the mass distribution by definition is hardly changing,
but to quickly become uncompetitive in the non-linear regime.

Lagrangian PP codes, on the other hand, are fully adaptive and can follow
arbitrary changes to the mass distribution.  However, the cost penalty for
not using a grid is so high that PP methods should be regarded as methods of
last resort.  In fact, it would be worth considerable effort to develop
hybrid or overlapping grids (or expansion centers) when the problem warrants,
such as have been developed by Villumsen (1983), or the almost completed
effort by Weeks (1988), for binary galaxies.

\section{Noise}

Particle noise is the major difficulty when modelling collisionless systems.
In a simulation with $N \sim 100$K, density fluctuations due to shot noise are
larger than those in a real elliptical galaxy, say, by factor of $\sim 1000$.
 This problem affects all spatial scales from the importance of two body
encounters to global variations.  The implication is therefore that we
should try to maximize $N$ in order to minimize the noise.  Such a strategy
places a still higher premium on computational efficiency.

\subsection{Smoothing or bias}

As simulations with sufficient spatial resolution to make meaningful use of
billions of particles are still not even remotely possible, some form of
smoothing of the density distribution is required.  However desirable
smoothing may seem, it must be borne in mind that it is a double-edged
sword that yields a {\it biased\/} estimate of the potential.

\begin{figure}[t]
\centerline{\psfig{figure=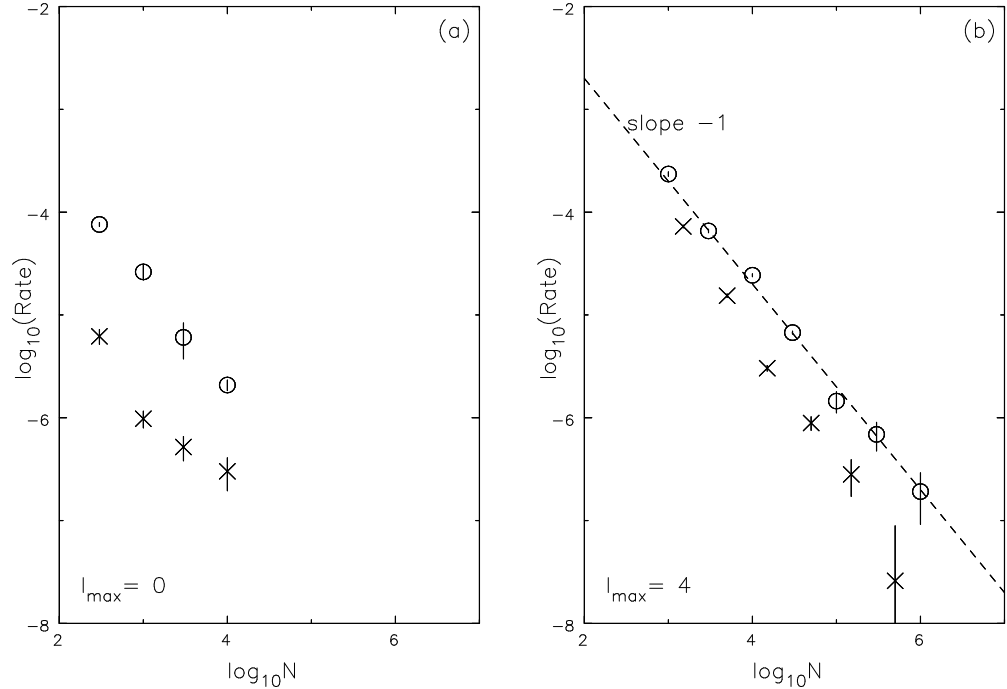,width=1.0\hsize,angle=270}}
\caption{The relaxation rate, defined in equation (2), for various numbers of
particles (a) when all non-radial terms are omitted and (b) when terms up to
$l_{\rm max}=4$ are included.  The dashed line is not a fit, but merely
indicates a of slope $-1$.  The circles are from experiments with randomly
distributed particles and the crosses are from experiments with quiet starts.}
\end{figure}

The nature of the bias is to limit resolution to the (effective) particle
size in PP and PM codes, which smooth only locally, and to leave larger
scales unsmoothed.  SFP methods, on the other hand, smooth globally but
thereby introduce a much more severe form of bias, since density changes that
are not represented by the truncated basis have no effect on the potential.
The ``{\it over}-smoothing'' strategy advocated by Weinberg (1996) reduces
the relaxation rate at the cost of further exacerbating the bias.

\subsection{Quiet starts}

A quiet start (Sellwood 1983) is a strategy for reducing noise without
increasing $N$.  It is achieved by distributing the particles smoothly
instead of at random.  The use of such techniques in astronomy dates back at
least to H\'enon (1968) and probably still further in plasma physics.  They
have become standard in laying down the initial conditions and perturbation
spectrum in cosmological simulations, as first clearly explained by
Doroshkevich et al.\ (1980), but continue to be largely ignored for galaxies.

Figure 3 shows empirical relaxation rates, defined by $$
\hbox{Rate} = {d \over dt} \left\langle \left[E_i(t) - E_i(0)\right]^2
\right\rangle, \eqno(2)
$$ where $E_i(t)$ is the specific energy of the $i$-th particle at time $t$
and the angle brackets indicate an average over all particles.  In a
perfectly collisionless simulation of a stable equilibrium model, each
particle would conserve its energy and this rate should be zero.  If the
relaxation is caused by shot noise in the particle distribution, we expect
${d \over dt}(\Delta E_{\rm rms}) \propto 1/\sqrt N$; the mean square changes
plotted seem consistent with a slope of $-1$.

Once again, the data are taken from simulations of the Plummer sphere
computed, in every case, using the PM+SH method with 201 radial grid points
and $l_{\rm max}=0$ or 4.  Simulations with particles placed at random
initially (noisy starts) follow the expected $1/N$ behavior.  Models begun
with a quiet start, however, relax more slowly -- apparently at about the
rate for a noisy start with several times the number of particles.

As pointed out by Weinberg (1993), collective neutral oscillations of the
system excited by the random distribution of particles are the most important
source of relaxation.  In the case of the Plummer sphere, experiments with
purely radial forces (Figure 3a) showed that relaxation is most effectively
reduced by suppressing collective radial pulsations of this spherical model.
The most important part of the quiet start strategy in this case is to space
several particles (10 in this case) having identical energy and angular
momentum components at equal intervals of radial phase.  This strategy
reduces the density variations caused by the radial oscillation of randomly
selected particles as they pursue their orbits, leading to an impressive
reduction in the relaxation rate.

Once non-radial forces are included, however, yet more particles are required
to reduce the non-radial density variations.  The data in Figure 3(b) were
obtained by placing replicas of each independent particle at 72$^\circ$
intervals around a circular ring that lies in the plane of that orbit (normal
to its angular momentum vector).  All the replica particles lie at a single
radius and have equal radial and azimuthal components of velocity and would
therefore follow exactly congruent orbits in that plane if the potential
were smooth.  This careful set up procedure leads to a relaxation rate that
seems to be roughly equivalent to that from a model having several times the
number of particles.

The number of particles per ring should exceed $l_{\rm max}$, five seem to be
adequate when $l_{\rm max}=4$.  The use of 10 rings of 5 particles to smooth
both the radial and azimuthal density variations reduces the number of
independent orbits by a factor of 50.  It is therefore desirable to take
extra care to select a better-than-random set of orbits ($E$ \& $L$ values)
from the \DF.

It should be noted that while it is easier to suppress noise by truncating
the angular expansion at low order, a similar procedure works well on a 3-D
Cartesian grid where no angular symmetries are imposed.  Again a smooth
radial distribution and ten particles per ring produced a very significant
reduction in the relaxation rate.

Provided the relaxation time greatly exceeds the duration of the experiment,
we can regard a simulation as adequately collisionless.  Thus in simulations
of mergers or collapses, which may need to be followed for only a few
dynamical times, relaxation caused by particle noise can be rendered
unimportant even with a modest number of randomly placed particles.  Quiet
starts are therefore advantageous only for initially equilibrium models that
are to be evolved in isolation for long periods.

\section{Conclusions}

While it remains true that no code is ``best'' for {\it every\/} problem, PM
codes have continued to hold their pre-eminent position as the most efficient
and should be considered first for many problems.  PM methods do have a
number of drawbacks, principally stemming from the finite volume enclosed by
the mesh, and therefore are not appropriate for every problem.  PP methods,
on the other hand, are so expensive they should be regarded as an option of
last resort for collisionless problems.

I have argued that the resolution limitations of PM methods are generally
overstated, since it is often possible to tailor the grid geometry to the
problem at hand.  Cartesian grids are not the only option, both cylindrical
and spherical grids have been used to excellent effect to improve resolution
dramatically near the centers.  The moral here is that if attaining
satisfactory resolution of the mass distribution of an isolated galaxy seems
to require an excessively large number of grid points, then the grid geometry
is probably inappropriate.  For problems I have worked on, making this change
has seemed more efficient than resorting to adaptive mesh refinement, though
this conclusion may not be always true.

My other principal conclusion is that it is rarely necessary to employ many
millions of particles for collisionless problems.  I find that $100\hbox{K}
\leq N \leq 500\hbox{K}$ is usually adequate and that results with more
particles are generally indistinguishable.  The relaxation rate, which
clearly should be much slower than the rate of evolution being simulated, can
be reduced without increasing $N$ simply by employing quiet starts.

With efficient PM codes and quiet starts, I find there is a wealth of
interesting problems that can be addressed quite adequately without requiring
supercomputers.

\bigskip
\noindent{\bf Acknowledgments}  I would like to thank Gerald Quinlan for
making his copies of NBODY1 and Hernquist's basis function codes available for
timing tests, and for his comments on a draft of this paper.  Also S.\ 
Shandarin pointed out the early reference to quiet starts in cosmology.  This 
work was supported by NSF grant AST 93/18617 and NASA Theory grant NAG 5-2803.

\end{document}